\documentclass[10pt, aps, prd,
reprint, % for double column
notitlepage, nofootinbib, longbibliography, showpacs, letter, eqsecnum,
superscriptaddress,
]{revtex4-1}

\usepackage{pstricks,pst-fractal}
\usepackage{amsmath,amssymb,bm}
\usepackage{graphicx}
\usepackage{verbatim}
\usepackage[breaklinks=true,colorlinks=true,urlcolor=blue,linkcolor=red,citecolor=red]{hyperref}

\begin{document}

\title{Casimir force for magnetodielectric media}

\author{Iver Brevik}
\email{iver.h.brevik@ntnu.no}
%\homepage{http://folk.ntnu.no/iverhb}
\affiliation{Department of Energy and Process Engineering,
Norwegian University of Science and Technology, NO-7491 Trondheim, Norway}

\author{Prachi Parashar}
\email{Prachi.Parashar@jalc.edu} 
%\homepage{https://www.ntnu.edu/employees/prachi.parashar}
\affiliation{John A. Logan College, Carterville, Illinois 62918, USA}
\affiliation{Department of Energy and Process Engineering,
Norwegian University of Science and Technology, NO-7491 Trondheim, Norway}

\author{K. V. Shajesh}
\email{kvshajesh@gmail.com}
%\homepage{http://www.physics.siu.edu/~shajesh}
\affiliation{Department of Physics, Southern Illinois University--Carbondale,
Carbondale, Illinois 62901, USA}
\affiliation{Department of Energy and Process Engineering,
Norwegian University of Science and Technology, NO-7491 Trondheim, Norway}

\date{\today}

%--------------------------------------------
\begin{abstract}
Boyer showed that a perfect electrically conducting slab repels
a perfect magnetically conducting slab, in contrast to the
attractive Casimir force between two identical perfect
electrically or magnetically conducting slabs. To gain insight for
the difference between the Boyer force and the Casimir force, we
present the derivation of the Boyer force using the stress tensor
method and then using the method of variation in dielectric.
The Green dyadic, in terms of electric
and magnetic Green's functions, is presented for an electric
medium filling half of space and another magnetic medium
filling another half of space such that the two half-spaces
are parallel and separated by a distance $a$. We make the
observation that the spectral distribution of scattering in a
Boyer cavity is that of Fermi-Dirac type, while the spectral
distribution of scattering in a Casimir cavity is that of
Bose-Einstein type. Based on this observation we conclude that
the difference between the Boyer force and the Casimir force
is governed by the statistics of the possible scattering in
the respective cavities.
\end{abstract}

\maketitle
%--------------------------------------------
%\tableofcontents
%--------------------------------------------

\section{Introduction}
\label{sec1}

\begin{figure}
\includegraphics{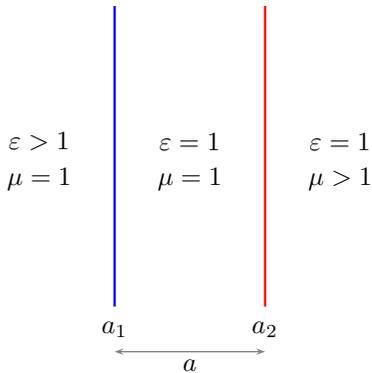}
\caption{Boyer configuration of parallel slabs.}
\label{fig-boyer-config}
\end{figure}

\begin{figure}
\includegraphics{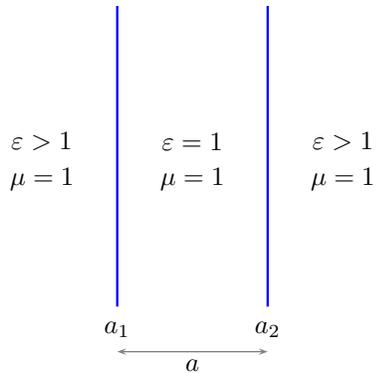}
\caption{Casimir configuration of parallel slabs.}
\label{fig-casimir-config}
\end{figure}

Consider the Boyer configuration of parallel slabs, consisting of
two parallel semi-infinitely thick plates, separated by a
vacuum gap $a$. Let the left plate be situated at $z=a_1$,
the right plate at $z=a_2$, so that $a=a_2-a_1$, 
see Fig.~\ref{fig-boyer-config}. Let the right plate
be purely paramagnetic ($\mu >1, \varepsilon=1$), and the left plate
purely dielectric ($\varepsilon >1, \mu=1$). 
The Boyer configuration of slabs should be contrasted with the
Casimir configuration of parallel slabs in Fig.~\ref{fig-casimir-config}. 
We shall be interested in the pressure, called $P$ here, between the slabs.
For simplicity we limit ourselves to zero temperature.
(General treatises on  the Casimir effect can be found in
Refs.~\cite{milton01,bordag09,dalvit11}.)

Stimulated by a suggestion from  Casimir, Boyer~\cite{boyer74}
made an explicit calculation of the force between an
infinitely permeable magnetic medium ($\mu \rightarrow \infty$)
and a perfect conductor ($\varepsilon \rightarrow \infty$).
The interaction energy per unit area was found to be
\begin{equation}
U=\frac{7}{8}\, \frac{\pi^2\hbar c}{720 a^3}, \label{1}
\end{equation}
which corresponds to a {\it repulsive} pressure
\begin{equation}
P=\frac{7}{8}\, |P_0|, \label{2}
\end{equation}
where
\begin{equation}
P_0=-\frac{\pi^2\hbar c}{240 a^4} \label{3}
\end{equation}
is the attractive force between two perfectly conducting
plates. As explained by Boyer, this is related to the fact that
while an electrically polarizable particle is attracted to a
conducting wall, a magnetically polarizable particle is repelled
by it (the latter case being  analogous to the hydrodynamic flow
from a  point or line source outside an impenetrable  plane).

From the standpoint of optical physics this result is however
rather perplexing. Consider for definiteness the electromagnetic
force density {\bf f} on the boundary layer in the left plate;
this force is purely electric, and is calculated from the
divergence of Maxwell's stress tensor to be
\begin{equation}
{\bf f}=-\frac{1}{2} E^2 {\bm\nabla} \varepsilon.  
\label{4}
\end{equation}
In classical physics this force always acts towards the optically
thinner medium, that is, towards the vacuum region for the plate
in consideration. One can observe that the {\it direction} of this
force given by the direction of ${\bm\nabla}\varepsilon$ is
independent of the magnetic properties of the right plate.
The direction is determined by the gradient of the dielectric
permittivity $\varepsilon$ only. Even if the magnetic properties
of the right plate should change, the electric field ${\bf E}$ on
the left plate would change, but the direction of this force would be
just the same.

The magnitude of the surface pressure $P$ can be found by
integrating the normal component $f_z$ across the boundary located at
$z=a_1$. Now  there are numerous cases in optics showing the reality
of the expression in Eq.\,(\ref{4}). For example, the classic
experiment of Ashkin and Dziedzic~\cite{ashkin73}, demonstrating
the outward bulge of a water surface illuminated by a radiation beam
coming from above, is of this sort, as is the newer version of this
experiment due to Astrath et al.~\cite{astrath14} (a review of some
radiation pressure experiments of this sort can be found in 
Ref.~\cite{brevik17}). Also, the recent  pressure experiment of
Kundu et al.~\cite{kundu17}, showing the deflection of a graphene
sheet upon laser illumination, belongs to the same category. In all
these cases, the force was found to act in the direction of the
optically thinner medium, in accordance with the expression in
Eq.\,(\ref{4}).

%---
%---

If one leaves the regime of classical physics and moves on to the
quantum mechanical calculation of the Casimir pressure,
one finds analogous results as long as both plates, separated
by a vacuum gap, are either {\it purely dielectric or purely magnetic}.
This pressure is conventionally calculated by taking the difference
between the $zz$ components of Maxwell's stress tensor on the
two sides, making use of the fluctuation-dissipation theorem when
constructing the two-point functions for the electric and magnetic
fields. Again, the pressure is found to be attractive, in
accordance with Eq.~(\ref{4}).

Now return to the magnetodielectric case considered by Boyer.
At first glance one would think that the repulsiveness of the
force as mentioned above is in direct conflict with Eq.~(\ref{4}).
What is the physical reason for this? Formally, the situation would
mean that one has to reverse the sign of the  quadratic quantity $E^2$
in Eq.\,(\ref{4}). In view of this bizarre situation, one would
think that a revisit of the fundamental assumptions behind
electromagnetic theory in matter is desirable. That was one
of the motivations behind the present paper.

We first present the Green function formalism in a general way
in Sec.~\ref{sec-formalism}, and provide explicit solutions
for the Green dyadic in Sec.~\ref{sec-Green-dyadic},
where the two plates are each allowed to possess arbitrary,
even frequency-dependent, values of $\mu$ and $\varepsilon$,
and calculate the surface force formally in Sec.~\ref{sc}.
Thereafter we specialize to the Boyer case.
We make the observation that the spectral distribution of
scattering in a Boyer cavity is that of Fermi-Dirac type,
while the spectral distribution of scattering in a Casimir
cavity is that of Bose-Einstein type. Based on this observation
we suggest that the difference between the Boyer force and the
Casimir force is statistics of the possible scattering in
the respective cavities.

Next, we inquire if the Boyer force is sensitive to a cutoff
in the frequency response of permeability $\mu$. Calculations of the
Boyer force have usually assumed $\mu$ to be a constant for all
frequencies. It is easy to see that this is a over-simplified model
that violates one of the fundamentals of macroscopic theory
(cf. also the discussion in Ref.~\cite{landau84}): Start from
the ``microscopic'' Maxwell equation (here in Heaviside-Lorentz units)
\begin{equation}
{\bm \nabla} \times {\bf h}= \rho {\bf v} 
+ \frac{\partial {\bf e}}{\partial t}, \label{5}
\end{equation}
where $\rho \bf{v}$ is the local current density and ${\bf h}$
and ${\bf e}$ are the local magnetic and electric fields.
Space averaging gives $\overline{\bf h}={\bf B}$ with ${\bf B}$
the magnetic induction, and $\overline{\bf e}={\bf E}$ 
with ${\bf E}$ the macroscopic electric field. Thus
\begin{equation}
{\bm \nabla} \times {\bf B}= \overline{\rho {\bf v}} 
+ \frac{\partial {\bf E}}{\partial t}. \label{6}
\end{equation}
Subtracting the Maxwell equation
${\bm\nabla} \times {\bf H}= \partial {\bf D}/\partial t$ we get
\begin{equation}
\overline{\rho {\bf v}} ={\bm \nabla} \times {\bf M}
+\frac{\partial {\bf P}}{\partial t}, \label{7}
\end{equation}
with ${\bf M}={\bf B}-{\bf H}$ and ${\bf P}={\bf D}-{\bf E}$.

%---
%---

Consider now the general definition of the magnetic moment
${\bf m}$ of a body,
\begin{equation}
{\bf m}=\frac{1}{2}\int {\bf r} \times \overline{\rho{\bf v}}\,dV. \label{8}
\end{equation}
This can be compared with the expression
\begin{equation}
\frac{1}{2}\int {\bf r}\times ({\bm\nabla} \times {\bf M})\,dV
= \int {\bf M}\,dV, \label{9}
\end{equation}
which is derived using vector manipulations observing that ${\bf M}=0$
in the vacuum region outside the body. Since ${\bf m}=\int {\bf M}\,dV$
it follows that we can put $\overline{\rho {\bf v}}$ equal to
${\bm\nabla} \times {\bf M}$. Comparing with Eq.~(\ref{7}) we
conclude that the consistency of macroscopic electrodynamics depends
on the possibility to neglect the $\partial {\bf P}/\partial t$ term.
This can be made concrete further by assuming that the body of linear size
$l$ is exposed to an electromagnetic wave of frequency $\omega$.
We shall insert speed of light $c$, momentarily, in expressions
in this section for the benefit of the reader.
Order of magnitude estimates give 
$\partial P/\partial t \sim \omega E \sim \omega^2lH/c$
(the relationship $E\sim \omega lH/c$ used in the last estimate coming
from the equation 
${\bm \nabla} \times {\bf E}=-\partial {\bf B}/\partial t$
and $E\sim H$). The first term on the right hand side of Eq.\,(\ref{7}) is 
${\bm \nabla} \times {\bf M}=\chi {\bm \nabla} \times {\bf H}$
where $\chi$ is the susceptibility, and is estimated to be of order
$c\chi H/l$. Thus, our consistency condition  reduces to
\begin{equation}
l^2 \ll \chi c^2/\omega^2; \label{10}
\end{equation}
an inequality already given in Ref.~\cite{landau84}.
It is easily seen that the inequality in Eq.\,(\ref{10}) is broken
already at frequencies $\omega$ much less than optical frequencies.
And that is confirmed by experiments also. For instance, for ferrite
with $\mu \approx 600$ the  maximum of the frequency $\omega/2\pi$
is reported to lie in the region 100\,kHz to 1\,MHz~\cite{NPL:KL}.
The assumption about a very large and constant permeability as used
in the Boyer calculation is obviously unphysical. 

In Sec.~\ref{sec-finitestm} we take the magnetic dispersion into account
in a crude way, by assuming $\mu$ to be constant up to a critical
value $\omega=\omega_c$. For higher frequencies we set $\mu=1$. Thus
\begin{equation}
\mu(\omega)=\begin{cases}
\mu, & \omega < \omega_c, \\
1, & \omega >\omega_c. \end{cases} \label{11}
\end{equation}
As one might expect, we do not find a reversal of the sign of
the force in this way; the force still comes out repulsive.
So the basic problem alluded to at the beginning, is not solved.
However, an important physical result of the calculation is that
the force turns out to be immeasurably small, when realistic input
data for $\omega_c$ are inserted. This makes it evident why the
repulsive force has never been measured.
It may finally be mentioned that we will keep $\varepsilon$ constant,
at all frequencies. There is no comparably strong limit to the
permittivity as it is to the permeability.

In Sec.~\ref{secintro} we consider the Boyer problem from a
statistical mechanical point of view, introducing a model where the
two media are represented by harmonic oscillators 1 and 2,
interacting with each other via a third oscillator 3. In this way
a mechanical analogue to the conventional TE and TM modes in
electromagnetism is obtained. We find that the quantum mechanical
transition from the TM to the TE mode, implying that oscillator 3
interacts with oscillators 1 and 2 via canonical momenta instead
of positions, is an important point. In this way the statistical
mechanical analysis helps us to elucidate the problem.

%--------------------------------------------
\section{Maxwell's equations}
\label{sec-formalism}

In Heaviside-Lorentz units the monochromatic components of Maxwell's equations,
proportional to $\exp(-i\omega t)$, in the absence of net charges and currents,
and in presence of dielectric and magnetic materials with boundaries, are
\begin{subequations}
\begin{align}
{\bm \nabla} \times {\bf E} &= i \omega {\bf B}, \label{MEcrossE} \\
-{\bm \nabla} \times {\bf H} &= i \omega ({\bf D} + {\bf P}).
\label{MEcrossB}
\end{align}%
\label{ME-cross}%
\end{subequations} 
These equations imply ${\bm \nabla} \cdot {\bf B}=0$ and
${\bm \nabla} \cdot ({\bf D}+ {\bf P})=0$.
Here ${\bf P}$ is an external source of polarization, in addition
to the polarization of the material in response to the fields 
${\bf E}$ and ${\bf H}$. The external source ${\bf P}$ serves as 
a convenient mathematical tool, and is set to zero in the end.
In the following we neglect non-linear responses and assume that 
the fields ${\bf D}$ and ${\bf B}$ respond linearly to the 
electric and magnetic fields ${\bf E}$ and ${\bf H}$:
\begin{subequations}
\begin{align}
{\bf D}({\bf r},\omega) 
&= {\bm \varepsilon}({\bf r};\omega) \cdot {\bf E}({\bf r},\omega), \\
{\bf B}({\bf r},\omega) 
&= {\bm \mu}({\bf r};\omega) \cdot {\bf H}({\bf r},\omega).
\end{align}%
\label{DB=emuEB}%
\end{subequations} 
Using Eq.\,(\ref{MEcrossE}) in Eq.\,(\ref{MEcrossB}) we construct the following
differential equation for the electric field,
\begin{equation}
\left[ \frac{1}{\omega^2} {\bm\nabla} \times {\bm\mu}^{-1} \cdot {\bm\nabla}
\times \;-{\bf 1} - {\bm\chi} \right] \cdot {\bf E}({\bf r},\omega)
= {\bf P}({\bf r},\omega),
\label{ddE=P}
\end{equation}
where 
\begin{equation}
{\bm\chi}({\bf r};\omega) = {\bm\varepsilon}({\bf r};\omega) - {\bf 1}.
\end{equation} 
The differential equation for the Green's dyadic 
${\bm\Gamma}({\bf r},{\bf r}^\prime;\omega)$ is guided by Eq.\,(\ref{ddE=P}),
\begin{equation}
\left[ \frac{1}{\omega^2} {\bm\nabla} \times {\bm\mu}^{-1} \cdot {\bm\nabla}
\times \;-{\bf 1} - {\bm\chi} \right] \cdot 
{\bm\Gamma}({\bf r},{\bf r}^\prime;\omega)
= {\bf 1} \delta^{(3)}({\bf r}-{\bf r}^\prime).
\label{Gd-deq}
\end{equation}
It defines the relation between the electric field and the polarization source,
\begin{equation}
{\bf E}({\bf r};\omega) 
= \int d^3r^\prime {\bm\Gamma}({\bf r},{\bf r}^\prime;\omega)
\cdot {\bf P}({\bf r}^\prime;\omega).
\label{E=GP-def}
\end{equation}
The corresponding dyadic for vacuum, obtained by setting ${\bm\chi}=0$, 
is called the free Green's dyadic and satisfies the equation
\begin{equation}
\left[ \frac{1}{\omega^2} {\bm\nabla} \times {\bm\mu}^{-1} \cdot {\bm\nabla}
\times \;-{\bf 1} \right] \cdot {\bm\Gamma}_0({\bf r},{\bf r}^\prime;\omega)
= {\bf 1} \delta^{(3)}({\bf r}-{\bf r}^\prime).
\label{fGd-deq}
\end{equation}
We can also define the relation between the magnetic field 
and the polarization source,
\begin{equation}
{\bf H}({\bf r};\omega) 
= \int d^3r^\prime {\bm\Phi}({\bf r},{\bf r}^\prime;\omega)
\cdot {\bf P}({\bf r}^\prime;\omega).
\label{H=GP-def}
\end{equation}

%-----------------------------------------------------

\subsection{Quantum electrodynamics}

The Green dyadics gives the correlation between the fields and sources,
as per Eqs.\,(\ref{E=GP-def}) and (\ref{H=GP-def}),
\begin{subequations}
\begin{align}
\frac{\delta {\bf E}({\bf r};\omega)}{\delta {\bf P}({\bf r}^\prime;\omega)}
&= {\bm\Gamma}({\bf r},{\bf r}^\prime;\omega), \\
\frac{\delta {\bf H}({\bf r};\omega)}{\delta {\bf P}({\bf r}^\prime;\omega)}
&= {\bm\Phi}({\bf r},{\bf r}^\prime;\omega).
\end{align}
\end{subequations}
In quantum electrodynamics the Green dyadics also serve as correlations
between the fields at two different points in space, which are stated as
\begin{subequations}
\begin{align}
\frac{1}{\tau}
\langle {\bf E}({\bf r};-\omega) {\bf E}({\bf r}^\prime;\omega) \rangle
&= \frac{1}{i} {\bm\Gamma}({\bf r},{\bf r}^\prime;\omega), \label{EE=iG} \\
\frac{1}{\tau}
\langle {\bf H}({\bf r};-\omega) {\bf H}({\bf r}^\prime;\omega) \rangle
&= \frac{1}{i} {\bm\Gamma}({\bf r},{\bf r}^\prime;\omega)
\Big|_{E\leftrightarrow H, \varepsilon \leftrightarrow \mu}, \\
\frac{1}{\tau}
\langle {\bf H}({\bf r};-\omega) {\bf E}({\bf r}^\prime;\omega) \rangle
&= \frac{1}{i} {\bm\Phi}({\bf r},{\bf r}^\prime;\omega) \\
\frac{1}{\tau}
\langle {\bf E}({\bf r};-\omega) {\bf H}({\bf r}^\prime;\omega) \rangle^*
&= \frac{1}{i} {\bm\Phi}({\bf r},{\bf r}^\prime;\omega),
\end{align}%
\label{cor-fields-GD}%
\end{subequations}%
where $\tau$ is the average (infinite) time for which the system
is observed.

%-----------------------------------------------------
\section{Green's dyadic}
\label{sec-Green-dyadic}

For planar geometry, using translational symmetry in the plane,
we can define the Fourier transformations
\begin{subequations}
\begin{align}
{\bm\Gamma}({\bf r},{\bf r}^\prime;\omega) &= \int \frac{d^2k_\perp}{(2\pi)^2} 
\,e^{i{\bf k}_\perp \cdot ({\bf r}-{\bf r}^\prime)_\perp} 
{\bm\gamma}(z,z^\prime;k_\perp,\omega), \\
{\bm\Phi}({\bf r},{\bf r}^\prime;\omega) &= \int \frac{d^2k_\perp}{(2\pi)^2} 
\,e^{i{\bf k}_\perp \cdot ({\bf r}-{\bf r}^\prime)_\perp} 
{\bm\phi}(z,z^\prime;k_\perp,\omega).
\end{align}%
\label{Gdya-ft}%
\end{subequations}%
The reduced Green's dyadics can be expressed in the form
\begin{equation}
{\bm\gamma}%(z,z^\prime;k_\perp,\omega)
= \left[ \begin{array}{ccc}
\frac{1}{\varepsilon^\perp(z)} \frac{\partial}{\partial z}
\frac{1}{\varepsilon^\perp(z^\prime)} \frac{\partial}{\partial z^\prime}
 g^H & 0 &
\frac{1}{\varepsilon^\perp(z)} \frac{\partial}{\partial z}
\frac{ik_\perp}{\varepsilon^{||}(z^\prime)} g^H \\[2mm]
0 & \omega^2 g^E & 0 \\[2mm]
-\frac{ik_\perp}{\varepsilon^{||}(z)} \frac{1}{\varepsilon^\perp(z^\prime)} 
\frac{\partial}{\partial z^\prime} g^H & 0 &
-\frac{ik_\perp}{\varepsilon^{||}(z)}
\frac{ik_\perp}{\varepsilon^{||}(z^\prime)} g^H
\end{array} \right].
\label{Gamma=gE}
\end{equation}
Here we have omitted the term
\begin{equation}
-\delta(z-z^\prime)
\left[ \begin{array}{llr}
\frac{1}{\varepsilon^\perp(z)} \hspace{2mm}& 0 \hspace{2mm} & 0 \\
0 & 0 & 0 \\ 0 & 0 & \frac{1}{\varepsilon^{||}(z)} \end{array} \right]
\label{omdterm}
\end{equation}
that contains a $\delta$-function and thus never contributes to 
interaction energies between two bodies unless they are overlapping, 
and
\begin{equation}
{\bm\phi}%(z,z^\prime;k_\perp,\omega)
= i\omega \left[ \begin{array}{ccc}
0 & \frac{1}{\mu^\perp(z)} \frac{\partial}{\partial z} g^E
& 0 \\[2mm]
\frac{1}{\varepsilon^\perp(z^\prime)}
\frac{\partial}{\partial z^\prime} g^H & 0 &
\frac{ik_\perp}{\varepsilon^{||}(z^\prime)} g^H \\[2mm]
0 & -\frac{ik_\perp}{\mu^{||}(z)} g^E & 0
\end{array} \right].
\label{Phi=gH}
\end{equation}
Here the magnetic (TM mode) Green's function $g^H(z,z^\prime)$, and 
the electric (TE mode) Green's function $g^E(z,z^\prime)$,
are defined using the differential equations
\begin{subequations}
\begin{eqnarray}
\left[ - \frac{\partial}{\partial z} \frac{1}{\varepsilon^\perp(z)}
\frac{\partial}{\partial z} + \frac{k_\perp^2}{\varepsilon^{||}(z)} 
-\omega^2 \mu^\perp(z) \right] g^H%(z,z^\prime)
 &=& \delta(z-z^\prime), \hspace{9mm}
\label{greenH} \\
\left[ - \frac{\partial}{\partial z} \frac{1}{\mu^\perp(z)}
\frac{\partial}{\partial z} + \frac{k_\perp^2}{\mu^{||}(z)} 
-\omega^2 \varepsilon^\perp(z) \right] g^E%(z,z^\prime)
 &=& \delta(z-z^\prime),
\label{greenE}
\end{eqnarray}%
\label{green-funs}%
\end{subequations}
where $\varepsilon = \text{diag}
(\varepsilon^\perp,\varepsilon^\perp,\varepsilon^{||})$ is the 
permittivity tensor and
$\mu = \text{diag} (\mu^\perp,\mu^\perp,\mu^{||})$ is the 
permeability tensor~\cite{Parashar:2012pti}.

%--------------------------------------------
\subsection{Electric Green's function for Boyer configuration}

%---
\begin{figure}
\includegraphics{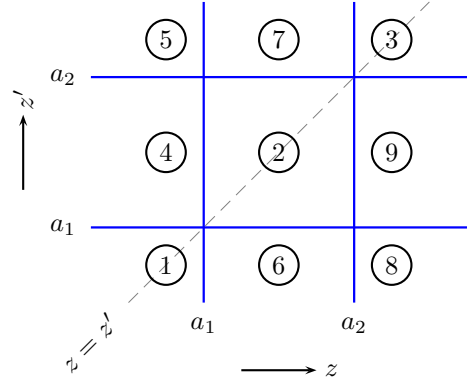}
\caption{Labels for the regions in the $z$-$z^\prime$ space,
used for specifying the Green function.}
\label{fig-regions-define}
\end{figure}
%---

For the Boyer configuration of slabs in Fig.~\ref{fig-boyer-config}
with isotropic permittivity $\varepsilon$ and permeability $\mu$
we have the differential equation for the electric Green's function
\begin{equation}
\left[ - \frac{\partial}{\partial z} \frac{1}{\mu(z)}
\frac{\partial}{\partial z} + \frac{k_\perp^2}{\mu(z)}
-\omega^2 \varepsilon(z) \right] g^E(z,z^\prime) = \delta(z-z^\prime),
\label{gE-iso-simple}%
\end{equation}%
where
\begin{subequations}
\begin{align}
\varepsilon(z) &= 1 +(\varepsilon-1) \,\theta(a_1-z), \\
\mu(z) &= 1 +(\mu-1) \,\theta(z-a_2),
\end{align}
\end{subequations}
with boundary conditions
\begin{subequations}
\begin{eqnarray}
g^E(z,z^\prime) \Big|^{a_i+\delta}_{a_i-\delta} &=& 0, \\
\left\{ \frac{1}{\mu(z)} \frac{\partial}{\partial z}
g^E(z,z^\prime) \right\} \bigg|^{a_i+\delta}_{a_i-\delta} &=& 0.
\end{eqnarray}
\end{subequations}
In terms of shorthand notations for typesetting
\begin{equation}
\kappa = \sqrt{k_\perp^2 +\zeta^2},
\end{equation}
and
\begin{equation}
\kappa_\varepsilon = \sqrt{k_\perp^2 +\zeta^2\varepsilon}, \qquad
\kappa_\mu = \sqrt{k_\perp^2 +\zeta^2\mu},
\end{equation}
and
\begin{equation}
\bar\kappa_\varepsilon = \frac{\kappa_\varepsilon}{\varepsilon}, \qquad 
\qquad \bar\kappa_\mu = \frac{\kappa_\mu}{\mu},
\end{equation}
and
\begin{subequations}
\begin{align}
r_\varepsilon =r_1^E
&= \frac{\kappa_\varepsilon-\kappa}{\kappa_\varepsilon+\kappa}, 
& r_2^E &= 0, \\ 
r_1^H &= 0, &
r_2^H &= \frac{\bar\kappa_\mu-\kappa}{\bar\kappa_\mu+\kappa} =r_\mu,
\end{align}
\end{subequations}
and
\begin{subequations}
\begin{align}
\Delta^E &= 1-r_\varepsilon \bar r_\mu e^{-2\kappa a}, \\
\Delta^H &= 1-\bar r_\varepsilon r_\mu e^{-2\kappa a},
\end{align}%
\label{delEH}%
\end{subequations}%
where $\bar r_\varepsilon$ is obtained by replacing 
$\kappa_\varepsilon \to \bar\kappa_\varepsilon$ in $r_\varepsilon$,
and, similarly, $\bar r_\mu$ is obtained by replacing 
$\kappa_\mu \to \bar\kappa_\mu$ in $r_\mu$.
In the following the label in the subscript represents the 
region in Fig.~\ref{fig-regions-define} in which the variables
$z$ and $z^\prime$ reside.
The solution for the electric Green's function is,
\begin{widetext}
\begin{subequations}
\begin{align}
g^E_{ \rput(0.1,0){\scriptstyle{1}}
\pscircle[linewidth=0.2pt](0.1,0.01){0.13} }
(z,z^\prime) &= \frac{1}{2\kappa_\varepsilon}
e^{-\kappa_\varepsilon |z-z^\prime|}
+\frac{1}{2\kappa_\varepsilon} \frac{1}{\Delta^E}
\Big[ r_\varepsilon -\bar r_\mu e^{-2\kappa a} \Big]
e^{-\kappa_\varepsilon |a_1-z|} e^{-\kappa_\varepsilon |a_1-z^\prime|},
&& z,z^\prime< a_1<a_2, \\
g^E_{ \rput(0.1,0){\scriptstyle{4}}
\pscircle[linewidth=0.2pt](0.1,0.01){0.13} }
(z,z^\prime) &= \frac{1}{\kappa_\varepsilon +\kappa} \frac{1}{\Delta^E}
e^{-\kappa_\varepsilon |a_1-z|} e^{-\kappa |z^\prime-a_1|}
-\frac{1}{\kappa_\varepsilon +\kappa} \frac{1}{\Delta^E}
\Big[ \bar r_\mu e^{-\kappa a} \Big]
e^{-\kappa_\varepsilon |a_1-z|} e^{-\kappa |a_2-z^\prime|},
&& z<a_1<z^\prime<a_2, \\
g^E_{ \rput(0.1,0){\scriptstyle{5}}
\pscircle[linewidth=0.2pt](0.1,0.01){0.13} }
(z,z^\prime) &= \frac{1}{2\kappa} \frac{1}{\Delta^E}
\Big[ \frac{2\kappa}{\kappa_\varepsilon +\kappa}
\frac{2\kappa}{\bar\kappa_\mu +\kappa} e^{-\kappa a} \Big]
e^{-\kappa_\varepsilon |a_1-z|} e^{-\kappa_\mu |z^\prime-a_2|},
&& z<a_1<a_2<z^\prime,
\end{align}
\end{subequations}
and
\begin{subequations}
\begin{align}
g^E_{ \rput(0.1,0){\scriptstyle{6}}
\pscircle[linewidth=0.2pt](0.1,0.01){0.13} }
(z,z^\prime) &= \frac{1}{\kappa_\varepsilon +\kappa} \frac{1}{\Delta^E}
e^{-\kappa |z-a_1|} e^{-\kappa_\varepsilon |a_1-z^\prime|}
-\frac{1}{\kappa_\varepsilon +\kappa} \frac{1}{\Delta^E}
\Big[ \bar r_\mu e^{-\kappa a} \Big]
e^{-\kappa |a_2-z|} e^{-\kappa_\varepsilon |a_1-z^\prime|},
\hspace{5mm} z^\prime<a_1<z<a_2, \\
g^E_{ \rput(0.1,0){\scriptstyle{2}}
\pscircle[linewidth=0.2pt](0.1,0.01){0.13} }
(z,z^\prime) &= \frac{1}{2\kappa} e^{-\kappa |z-z^\prime|}
-\frac{1}{2\kappa} \frac{1}{\Delta^E}
\Big[ r_\varepsilon \Big] e^{-\kappa |z-a_1|} e^{-\kappa |z^\prime-a_1|}
+\frac{1}{2\kappa} \frac{1}{\Delta^E} 
\Big[ r_\varepsilon \bar r_\mu e^{-\kappa a} \Big]
e^{-\kappa |z-a_1|} e^{-\kappa |a_2-z^\prime|}
\nonumber \\ & \hspace{4mm}
+\frac{1}{2\kappa} \frac{1}{\Delta^E}
\Big[ \bar r_\mu r_\varepsilon e^{-\kappa a} \Big]
e^{-\kappa |a_2-z|} e^{-\kappa |z^\prime-a_1|}
-\frac{1}{2\kappa} \frac{1}{\Delta^E}
\Big[ \bar r_\mu \Big]
e^{-\kappa |a_2-z|} e^{-\kappa |a_2-z^\prime|},
\hspace{7mm} a_1<z^\prime,z<a_2, \\
g^E_{ \rput(0.1,0){\scriptstyle{7}}
\pscircle[linewidth=0.2pt](0.1,0.01){0.13} }
(z,z^\prime) &= \frac{1}{\bar\kappa_\mu +\kappa} \frac{1}{\Delta^E}
e^{-\kappa |a_2-z|} e^{-\kappa_\mu |z^\prime-a_2|}
-\frac{1}{\bar\kappa_\mu +\kappa} \frac{1}{\Delta^E}
\Big[ r_\varepsilon e^{-\kappa a} \Big]
e^{-\kappa |z-a_1|} e^{-\kappa_\mu |z^\prime-a_2|},
\hspace{5mm} a_1<z<a_2<z^\prime,
\end{align}
\end{subequations}
and
\begin{subequations}
\begin{align}
g^E_{ \rput(0.1,0){\scriptstyle{8}}
\pscircle[linewidth=0.2pt](0.1,0.01){0.13} }
(z,z^\prime) &= \frac{1}{2\kappa} \frac{1}{\Delta^E}
\Big[ \frac{2\kappa}{\kappa_\varepsilon +\kappa}
\frac{2\kappa}{\bar\kappa_\mu +\kappa} e^{-\kappa a} \Big]
e^{-\kappa_\mu |z-a_2|} e^{-\kappa_\varepsilon |a_1-z^\prime|},
&& z^\prime<a_1<a_2<z, \\
g^E_{ \rput(0.1,0){\scriptstyle{9}}
\pscircle[linewidth=0.2pt](0.1,0.01){0.13} }
(z,z^\prime) &= \frac{1}{\bar\kappa_\mu +\kappa} \frac{1}{\Delta^E}
e^{-\kappa_\mu |z-a_2|} e^{-\kappa |a_2-z^\prime|}
-\frac{1}{\bar\kappa_\mu +\kappa} \frac{1}{\Delta^E}
\Big[ r_\varepsilon e^{-\kappa a} \Big]
e^{-\kappa_\mu |z-a_2|} e^{-\kappa |z^\prime-a_1|},
&& a_1<z^\prime<a_2<z, \\
g^E_{ \rput(0.1,0){\scriptstyle{3}}
\pscircle[linewidth=0.2pt](0.1,0.01){0.13} }
(z,z^\prime) &= \frac{1}{2\bar\kappa_\mu} e^{-\kappa_\mu |z-z^\prime|}
+\frac{1}{2\bar\kappa_\mu} \frac{1}{\Delta^E}
\Big[ \bar r_\mu -r_\varepsilon e^{-2\kappa a} \Big]
e^{-\kappa_\mu |z-a_2|} e^{-\kappa_\mu |z^\prime-a_2|},
&& a_1<a_2<z^\prime,z.
\end{align}
\end{subequations}
\end{widetext}

%--------------------------------------------
\subsection{Magnetic Green's function for Boyer configuration}

For the Boyer configuration of slabs in Fig.~\ref{fig-regions-define}
the magnetic Green's function has the differential equation
\begin{equation}
\left[ - \frac{\partial}{\partial z} \frac{1}{\varepsilon(z)}
\frac{\partial}{\partial z} + \frac{k_\perp^2}{\varepsilon(z)}
-\omega^2 \mu(z) \right] g^H(z,z^\prime) = \delta(z-z^\prime),
\label{gH-iso-simple}%
\end{equation}%
with boundary conditions
\begin{subequations}
\begin{eqnarray}
g^H(z,z^\prime) \Big|^{a_i+\delta}_{a_i-\delta} &=& 0, \\
\left\{ \frac{1}{\varepsilon(z)} \frac{\partial}{\partial z}
g^H(z,z^\prime) \right\} \bigg|^{a_i+\delta}_{a_i-\delta} &=& 0.
\end{eqnarray}
\end{subequations}
The solution for the magnetic Green's function is obtained from 
the electric Green's function
by replacing $\kappa_\varepsilon \to \bar\kappa_\varepsilon$ and 
$\bar\kappa_\mu \to \kappa_\mu$ everywhere, except in the exponentials.
This leads to the replacements,
$r_\varepsilon \to \bar r_\varepsilon$, $\bar r_\mu \to r_\mu$, and
$\Delta^E \to \Delta^H$.

%--------------------------------------------
\subsection{Perfect electric and magnetic conductors}
\label{sec-per-con-limit}

In the limit $\varepsilon\to\infty$ and $\mu\to\infty$,
when the region $z<a_1$ is a 
perfectly conducting electric medium and the region $z>a_1$
is a perfectly conducting magnetic material,
we observe that the region corresponding to $a_1<z,z^\prime<a_2$
in Fig.~\ref{fig-boyer-config} is
the only relevant region in the discussion as the fields vanish
inside the two perfectly conducting media.
Thus, we have
\begin{equation}
\Delta_\textrm{PC} = 1+ e^{-2\kappa a}.
\end{equation} 
In this case, the
explicit form for the Green functions can be conveniently expressed as 
\begin{subequations}
\begin{eqnarray}
g^E(z,z^\prime) &=& \frac{1}{\kappa}
\frac{\sinh\kappa (z_<-a_1) \cosh\kappa (z_>-a_2)}{\cosh\kappa a}, 
\hspace{11mm} \\
g^H(z,z^\prime) &=& -\frac{1}{\kappa}
\frac{\cosh\kappa (z_<-a_1) \sinh\kappa (z_>-a_2)}{\cosh\kappa a},
\end{eqnarray}
\end{subequations}
which is evaluated in region $\rput(0,0.1){
\rput(0.1,0){\scriptstyle{2}} \pscircle[linewidth=0.2pt](0.1,0.01){0.13} }$
\hspace{3pt}.
Here $z_<=\text{Min}(z,z^\prime)$ and $z_>=\text{Max}(z,z^\prime)$.
We make the observation that, on the boundaries of the perfectly
conducting slabs at $z=a_1$ and $z=a_2$ of the Boyer configuration we have
\begin{subequations}
\begin{align}
g^E(a_1,a_2) &=0, & g^E(a_2,a_2) &= \frac{\tanh\kappa a}{\kappa}, \\
g^E(a_1,a_1) &=0, & g^E(a_2,a_1) &= \frac{\sinh\kappa a}{\kappa},
\end{align}
\end{subequations}
and
\begin{subequations}
\begin{align}
g^H(a_1,a_2) &=0, & g^H(a_2,a_2) &= 0, \\
g^H(a_1,a_1) &= \frac{\tanh\kappa a}{\kappa}, & 
g^H(a_2,a_1) &= \frac{\sinh\kappa a}{\kappa}.
\end{align}
\end{subequations}
We further make the observation that
\begin{subequations}
\begin{eqnarray}
\frac{\partial}{\partial z} \frac{\partial}{\partial z^\prime}
g^E(z,z^\prime) &=& -\kappa^2 g^H(z,z^\prime), \\
\frac{\partial}{\partial z} \frac{\partial}{\partial z^\prime}
g^H(z,z^\prime) &=& -\kappa^2 g^E(z,z^\prime), 
\end{eqnarray}
\end{subequations}
such that
\begin{eqnarray}
\frac{\partial}{\partial z} \frac{\partial}{\partial z^\prime}
\Big[ g^E(z,z^\prime) + g^H(z,z^\prime) \Big]
\bigg|_{z=a_1,z^\prime=a_1} \hspace{15mm} \nonumber \\
=-\kappa^2 \Big[ g^E(z,z^\prime) + g^H(z,z^\prime) \Big], \hspace{10mm}
\end{eqnarray}
and
\begin{equation}
g^E(a_1,a_1) + g^H(a_1,a_1) = \frac{\tanh\kappa a}{\kappa}.
\label{gegHtan}
\end{equation}

%--------------------------------------------
\section{Strong coupling: Boyer's result}
\label{sc}

For slowly varying fields, and assuming that the dissipation in
the system is negligible, the statement of conservation of momentum density,
\begin{equation}
\rho {\bf E} + \rho {\bf v}\times {\bf B}
+\frac{\partial {\bf G}}{\partial t} +{\bm\nabla}\cdot {\bf T} 
-\frac{1}{2} E^2 {\bm\nabla} \varepsilon -\frac{1}{2} H^2 {\bm\nabla} \mu =0,
\label{for-def}
\end{equation}
can be derived starting from the Maxwell equations~\cite{Parashar:2018pds},
where
\begin{equation}
{\bf G} = {\bf D} \times {\bf B}
\end{equation}
is the momentum density of the electromagnetic field and
\begin{equation}
{\bf T} = {\bf 1} \frac{1}{2} ({\bf D} \cdot {\bf E} + {\bf B} \cdot {\bf H})
- ({\bf D} {\bf E} + {\bf B} {\bf H})
\label{stten-def}
\end{equation}
is the stress tensor or the flux of the momentum density
of the electromagnetic field.
Each term in Eq.\,(\ref{for-def}) is interpreted as a force when
it is integrated over a volume $V$.
The first two terms in Eq.\,(\ref{for-def}), together, contribute to the
Lorentz force acting on the charges inside volume $V$, 
\begin{equation}
{\bf f}_\text{Lor} = \rho {\bf E} + \rho {\bf v}\times {\bf B},
\end{equation}
where $\rho$ is the density of charge and ${\bf v}$ is the velocity of charges
that contributes to current density $\rho {\bf v}$. The Lorentz force
is zero for a neutral medium. The third term in Eq.\,(\ref{for-def}),
when integrated over volume $V$, is the measure of the rate of change
of electromagnetic momentum inside the volume $V$,
\begin{equation}
{\bf f}_m = \frac{\partial {\bf G}}{\partial t}.
\end{equation}
This term is zero for slowly varying fields.
The fourth term in Eq.\,(\ref{for-def}) measures the flux of the 
electromagnetic field across the surface of the volume $V$ and
when integrated over a volume $V$, as a consequence of the Gauss law,
measures the stress on the surface of the volume $V$ due to 
the electromagnetic field. The force density due to this stress
from the radiation is 
\begin{equation}
{\bf f}_\text{rad} = -{\bm\nabla}\cdot {\bf T}. 
\label{for-rad}
\end{equation}
The fifth and the sixth term in Eq.\,(\ref{for-def}) are the rate of
transfer of electromagnetic energy to the dielectric and permeable material
inside the volume $V$,
\begin{subequations}
\begin{eqnarray}
{\bf f}_\varepsilon &=&-\frac{1}{2} E^2 {\bm\nabla} \varepsilon,
\label{elfor-def} \\
{\bf f}_\mu &=& -\frac{1}{2} H^2 {\bm\nabla} \mu,
\end{eqnarray}
\end{subequations}
respectively. Together, we have
\begin{equation}
{\bf f}_\text{Lor} +{\bf f}_m -{\bf f}_\text{rad} 
+{\bf f}_\varepsilon +{\bf f}_\mu = 0.
\label{sum-force}
\end{equation}

%-----------------------------
\subsection{Stress tensor method}

%---
\begin{figure}
\includegraphics{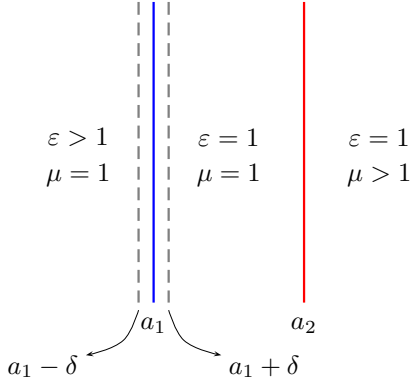}
\caption{Boyer configuration of parallel slabs
with the illustration of the integration volume $V$.
The integration volume $V$ represents an infinitely thin film
that encloses the surface of the dielectric slab at $z=a_1$.
The surfaces of the film are at $z=a_1-\delta$
and $z=a_1+\delta$, and take the limit $\delta\to 0$.}
\label{fig-boyer-intV}
\end{figure}
%---

For a neutral dielectric medium,
if we restrict to essentially static cases when the fields
are varying slowly, three of the five terms in Eq.\,(\ref{sum-force})
can be neglected. In this case we have the relation
\begin{equation}
{\bf f}_\text{rad} ={\bf f}_\varepsilon +{\bf f}_\mu.
\label{rad=eps-for}
\end{equation}
We choose the volume $V$ in our discussion to be an infinitely
thin film that encloses the surface of the dielectric medium.
In the Boyer configuration of Fig.~\ref{fig-boyer-config}
this has been illustrated in Fig.~\ref{fig-boyer-intV}.
Eq.\,(\ref{rad=eps-for}) is the statement of balance of forces.
Integrating Eq.\,(\ref{for-rad}) over the volume $V$,
and using Gauss's law, the left hand side of Eq.\,(\ref{rad=eps-for})
gives the total radiation force on the left plate,
\begin{equation}
{\bf F}_\text{rad}(t) = -\oint_V d{\bf S} \cdot {\bf T}({\bf r}, t).
\end{equation}
The time averaged radiation force ${\bf F}_\text{rad}$ is defined as
the time average of ${\bf F}_\text{rad}(t)$, $\tau =2T$,
\begin{equation}
{\bf F}_\text{rad} = \overline{{\bf F}_\text{rad}(t)}
= \frac{1}{\tau} \int_{-T}^T dt \,{\bf F}_\text{rad}(t). 
\end{equation}
The stress tensor is a bilinear construction of the fields.
For example, it involves the construction 
\begin{eqnarray}
{\bf E}({\bf r},t) {\bf D}({\bf r},t) \hspace{60mm} \nonumber \\
= \int_{-\infty}^\infty \frac{d\omega}{2\pi}
\int_{-\infty}^\infty \frac{d\omega^\prime}{2\pi}
e^{-i(\omega^\prime -\omega)t}
{\bf E}({\bf r},\omega)^* {\bf D}({\bf r},\omega^\prime), \hspace{8mm}
\end{eqnarray}
which uses the Fourier transformation
\begin{equation}
{\bf E}({\bf r},t) = \int_{-\infty}^\infty \frac{d\omega}{2\pi}
e^{-i\omega t} {\bf E}({\bf r},\omega).
\end{equation}
Time average of a bilinear construction satisfies the Plancherel theorem
\begin{equation}
\int_{-\infty}^\infty dt\, {\bf E}({\bf r},t) {\bf D}({\bf r},t)
= \int_{-\infty}^\infty \frac{d\omega}{2\pi}
{\bf E}({\bf r},\omega)^* {\bf D}({\bf r},\omega),
\end{equation}
which implies
\begin{equation}
\int_{-\infty}^\infty dt\, {\bf T}({\bf r},t)
= \int_{-\infty}^\infty \frac{d\omega}{2\pi} {\bf T}({\bf r},\omega).
\end{equation}
Thus, we have, presuming $\tau=2T\to\infty$,
\begin{subequations}
\begin{eqnarray}
{\bf F}_\text{rad}
&=& -\frac{1}{\tau} \int_{-\infty}^\infty dt 
\oint_V d{\bf S} \cdot {\bf T}({\bf r}, t) \\
&=& -\frac{1}{\tau} \int_{-\infty}^\infty \frac{d\omega}{2\pi} 
\oint_V d{\bf S} \cdot {\bf T}({\bf r}, \omega).
\end{eqnarray}
\end{subequations}

The fluctuations in the quantum vacuum do not contribute to the mean 
value of the field and thus the field satisfies
\begin{equation}
\langle {\bf E} \rangle =0,
\end{equation}
but they contribute non-zero correlations in bilinear constructions
of fields, and are contained in Eqs.\,(\ref{cor-fields-GD}).
In particular, quantum vacuum fluctuations leads to non-zero contributions
in the flux tensor. The radiation force arising from these
contributions in the flux tensor, that are manifestations of
the quantum vacuum, is a force given by
\begin{equation}
{\bf F}
= -\frac{1}{2T} \int_{-\infty}^\infty \frac{d\omega}{2\pi}
\oint_V d{\bf S} \cdot \langle {\bf T}({\bf r}, \omega) \rangle.
\label{casfor-def}
\end{equation}
We can write the force on the half-slab at $z=a_1$ to be 
\begin{equation}
\frac{{\bf F} \cdot \hat{\bf z}}{A} 
= \frac{1}{i} \int_{-\infty}^{\infty} \frac{d\zeta}{2\pi}
\int \frac{d^2k_\perp}{(2\pi)^2}\left[ T_{33}(a_1+\delta) -T_{33}(a_1-\delta)\right], 
\label{P-a1}
\end{equation}
where, using Eq.\,(\ref{stten-def}),
\begin{subequations}
\begin{eqnarray}
T_{33}(a_1-\delta) &=& \frac{1}{2}\Big[E_1^2 +E_2^2 -E_3^2\Big]
\bigg|_{a_1-\delta} \nonumber \\ &&
+ \frac{1}{2}\Big[H_1^2 +H_2^2 -H_3^2 \Big] \bigg|_{a_1-\delta} \\
&=& \frac{\varepsilon}{2i} \Big[\frac{\partial}{\varepsilon}
\frac{\partial^\prime}{\varepsilon}-\bar{\kappa}_\varepsilon^2\Big] 
g^H_{ \rput(0.1,0){\scriptstyle{1}}
\pscircle[linewidth=0.2pt](0.1,0.01){0.13} } (a_1,a_1)
\nonumber \\ &&
+ \frac{1}{2i} \Big[ \partial \partial^\prime-\kappa_\varepsilon^2\Big] 
g^E_{ \rput(0.1,0){\scriptstyle{1}}
\pscircle[linewidth=0.2pt](0.1,0.01){0.13} } (a_1,a_1) \hspace{15mm}
\label{t33-}
\end{eqnarray}
\end{subequations}
and
\begin{subequations}
\begin{eqnarray}
T_{33}(a_1+\delta) &=& \frac{1}{2}\Big[E_1^2 +E_2^2 -E_3^2\Big]
\bigg|_{a_1+\delta} \nonumber \\ && 
+ \frac{1}{2}\Big[H_1^2 +H_2^2 -H_3^2 \Big] \bigg|_{a_1+\delta}
\\ &=& -\frac{\kappa^2}{i}
\Big[ g^E_{ \rput(0.1,0){\scriptstyle{2}}
\pscircle[linewidth=0.2pt](0.1,0.01){0.13} } (a_1,a_1)
+ g^H_{ \rput(0.1,0){\scriptstyle{2}}
\pscircle[linewidth=0.2pt](0.1,0.01){0.13} } (a_1,a_1) \Big]. \hspace{8mm}
\label{t33+}
\end{eqnarray}
\end{subequations}
The stress tensor $T_{33}(a_1-\delta)$ is zero inside the perfect conductor.
Thus, using Eq.\,(\ref{gegHtan}), we obtain
\begin{equation}
P=\frac{{\bf F} \cdot \hat{\bf z}}{A}
= \frac{1}{2\pi^2} \int_0^\infty \kappa^3 d\kappa \,\tanh\kappa a,
\end{equation}
which can be rewritten in the form
\begin{equation}
P = \frac{1}{2\pi^2} \int_0^\infty \kappa^3 d\kappa
\left[ 1 - \frac{2}{e^{2\kappa a}+1} \right],
\end{equation}
in which we have separated the divergent bulk contribution that does
not have any information about the slabs.
Subtracting the bulk contribution we have
\begin{equation}
P= -\frac{1}{\pi^2} \int_0^\infty \frac{\kappa^3d\kappa}{e^{2\kappa a}+1}
= -\frac{7}{8} \frac{\pi^2}{240 a^4},
\label{boyer}
\end{equation}
which is exactly the result obtained by Boyer.
A negative force on the left plate corresponds to repulsion between the slabs.

%-----------------------------
\subsection{Variation in dielectric method}

The force on the dielectric slab can also be evaluated using the
force density on the right hand side of Eq.\,(\ref{rad=eps-for}),
which is given by Eq.\,(\ref{elfor-def}) and is expressed in a
more explicit form here,
\begin{equation}
{\bf f}_\varepsilon ({\bf r},t) 
= -\frac{1}{2} \int_{-\infty}^\infty dt^\prime
\Big[ {\bm\nabla} \varepsilon({\bf r},t-t^\prime) \Big]
\,{\bf E}({\bf r},t) \cdot {\bf E}({\bf r},t^\prime).
\end{equation}
In terms of frequency this takes the form
\begin{eqnarray}
{\bf f}_\varepsilon ({\bf r},t) 
&=& -\frac{1}{2} \int_{-\infty}^\infty \frac{d\omega}{2\pi} 
\int_{-\infty}^\infty \frac{d\omega^\prime}{2\pi} 
e^{-i(\omega-\omega^\prime)}
\nonumber \\ && \times
\Big[ {\bm\nabla} \varepsilon({\bf r},\omega) \Big]
\,{\bf E}({\bf r},\omega^\prime)^* \cdot {\bf E}({\bf r},\omega). \hspace{10mm} 
\end{eqnarray}
Time average of this force is calculated as, using $\tau=2T\to\infty$,
\begin{subequations}
\begin{eqnarray}
{\bf F}_\varepsilon ({\bf r}) = \frac{1}{\tau} 
\int_{-\infty}^\infty dt\, {\bf f}_\varepsilon ({\bf r},t) \hspace{45mm} \\
= -\frac{1}{2} \int_{-\infty}^\infty \frac{d\omega}{2\pi}
\Big[ {\bm\nabla} \varepsilon({\bf r},\omega) \Big] \frac{1}{\tau}
\,{\bf E}({\bf r},\omega^\prime)^* \cdot {\bf E}({\bf r},\omega). 
\hspace{11mm}
\label{forden-ve}
\end{eqnarray}
\end{subequations}
The force density,
using Eq.\,(\ref{EE=iG}) in Eq.\,(\ref{forden-ve}), is defined as
\begin{subequations}
\begin{eqnarray}
{\bf F}({\bf r}) 
&=& \langle {\bf F}_\varepsilon ({\bf r}) \rangle \\
&=& -\frac{1}{2} \int_{-\infty}^\infty \frac{d\omega}{2\pi}
\Big[ {\bm\nabla} \varepsilon({\bf r},\omega) \Big]
\,\text{tr}\, {\bm\Gamma} ({\bf r},{\bf r};\omega), \hspace{10mm}
\end{eqnarray}
\end{subequations}
where trace $\text{tr}$ is over the matrix index.
The total force on a volume $V$ is 
\begin{subequations}
\begin{eqnarray}
{\bf F} &=& \int_V d^3r\, {\bf F}({\bf r}) \\
&=& -\frac{1}{2} \int_{-\infty}^\infty \frac{d\omega}{2\pi} \int_V d^3r\,
\Big[ {\bm\nabla} \varepsilon({\bf r},\omega) \Big]
\,\text{tr}\, {\bm\Gamma} ({\bf r},{\bf r};\omega). \hspace{11mm}
\end{eqnarray}
\end{subequations}
This expression, in contrast to the expression for the 
Casimir force Eq.\,(\ref{casfor-def}),
is another expression for the Casimir force of a fundamentally 
different origin.
For planar configurations, using Eq.\,(\ref{Gdya-ft}),
and after switching to imaginary frequency,
$\omega \to i\zeta$, the Casimir force per unit area
on the slab at $z=a_1$ in Fig.~\ref{fig-boyer-config} is 
\begin{eqnarray}
\frac{{\bf F} \cdot \hat{\bf z}}{A} 
&=& -\frac{1}{2} \int_{-\infty}^\infty \frac{d\zeta}{2\pi}
\int \frac{d^2k_\perp}{(2\pi)^2}
\int_{a_1-\delta}^{a_1+\delta} dz\, 
\nonumber \\ && \times
\Big[ \hat{\bf z} \cdot {\bm\nabla} \varepsilon({\bf r},\omega) \Big]
\,\text{tr}\, {\bm\gamma} (z,z;{\bf k}_\perp,\omega). \hspace{10mm}
\end{eqnarray}
For the Boyer configuration in Fig.~\ref{fig-boyer-config} we have
\begin{equation}
\hat{\bf z} \cdot {\bm\nabla} \varepsilon({\bf r},\omega)
= -(\varepsilon-1) \delta(z-a_1).
\end{equation}
Assuming frequency independent dielectric function we have
\begin{equation}
\frac{{\bf F} \cdot \hat{\bf z}}{A}
= \frac{(\varepsilon-1)}{2} \int_{-\infty}^\infty \frac{d\zeta}{2\pi}
\int \frac{d^2k_\perp}{(2\pi)^2}
\,\text{tr}\, {\bm\gamma} (a_1,a_1;{\bf k}_\perp,\omega),
\label{cfintg}
\end{equation}
where
\begin{eqnarray}
\,\text{tr}\, {\bm\gamma} (a_1,a_1;{\bf k}_\perp,\omega)
= \frac{1}{\varepsilon(z)} \frac{\partial}{\partial z}
\frac{1}{\varepsilon(z^\prime)} \frac{\partial}{\partial z^\prime}
 g^H(z,z^\prime) \bigg|_{z=a_1,\,z^\prime=a_1} \nonumber \\
-\zeta^2 g^E(a_1,a_1) 
+\frac{k_\perp^2}{\varepsilon(z) \varepsilon(z^\prime)} g^H(a_1,a_1).
\hspace{20mm}
\end{eqnarray}
The continuity conditions for the Green functions allows the evaluation
of $g^E(a_1,a_1)$, $g^H(a_1,a_1)$, and
$\partial\partial^\prime g^H(a_1,a_1)/(\varepsilon\varepsilon^\prime)$,
without caring about from which region in Fig.~\ref{fig-regions-define}
the variables $z$ and $z^\prime$ approach $a_1$.
This still leaves the regions from which $z$ and $z^\prime$ approach $a_1$ in
\begin{equation}
\frac{k_\perp^2}{\varepsilon(z) \varepsilon(z^\prime)}
\end{equation}
undetermined. Following the suggestion in Ref.~\cite{Schwinger:1978dec}
we will require $z$ and $z^\prime$ to approach the interface at $a_1$
from opposite sides. 
This is necessary because otherwise the contribution to the Green
dyadic from Eq.\,(\ref{omdterm}) will contribute spurious divergences.
This is achieved by evaluating 
${\bm\gamma} (a_1,a_1;{\bf k}_\perp,\omega)$
in region $\rput(0,0.1){
\rput(0.1,0){\scriptstyle{4}} \pscircle[linewidth=0.2pt](0.1,0.01){0.13} }$
\hspace{3pt} or
in region $\rput(0,0.1){
\rput(0.1,0){\scriptstyle{6}} \pscircle[linewidth=0.2pt](0.1,0.01){0.13} }$
\hspace{3pt} 
of Fig.~\ref{fig-regions-define}. Thus,
\begin{eqnarray}
\,\text{tr}\, {\bm\gamma}^{\hspace{5pt}}_{ 
\rput(0.1,0){\scriptstyle{4}} \pscircle[linewidth=0.2pt](0.1,0.01){0.13} } 
(a_1,a_1;{\bf k}_\perp,\omega)
&=& \frac{1}{\varepsilon} \partial \partial^\prime
g^H_{\rput(0.1,0){\scriptstyle{4}} \pscircle[linewidth=0.2pt](0.1,0.01){0.13}} 
(a_1,a_1) +\frac{k_\perp^2}{\varepsilon}
g^H_{\rput(0.1,0){\scriptstyle{4}} \pscircle[linewidth=0.2pt](0.1,0.01){0.13}}
(a_1,a_1)  \nonumber \\ &&
-\zeta^2 
g^E_{\rput(0.1,0){\scriptstyle{4}} \pscircle[linewidth=0.2pt](0.1,0.01){0.13}} 
(a_1,a_1). 
\end{eqnarray}
The pressure in Eq.\,(\ref{cfintg}) involves the evaluation of
\begin{equation}
(\varepsilon-1)\,\text{tr}\, {\bm\gamma}^{\hspace{5pt}}_{ 
\rput(0.1,0){\scriptstyle{4}} \pscircle[linewidth=0.2pt](0.1,0.01){0.13} } 
(a_1,a_1;{\bf k}_\perp,\omega),
\label{trgepEv}
\end{equation}
which needs a little care when taking the perfect conductor limit. Observe
that the electric Green function $g^E(a_1,a_1)=0$. Thus, the term
$\zeta^2(\varepsilon-1) g^E(a_1,a_1)$ in the perfect conductor limit
is ambiguous if we take the perfect conductor limit early in the calculation.
To this end we define the parameter
\begin{equation}
s = 1-r_\varepsilon,
\end{equation}
which goes to zero in the perfect electric conductor limit. We note that
\begin{eqnarray}
g^E_{\rput(0.1,0){\scriptstyle{4}} \pscircle[linewidth=0.2pt](0.1,0.01){0.13}}
(a_1,a_1) = \frac{s}{2\kappa}
\left[ 1 - s \frac{\bar r_\mu e^{-2\kappa a}}{\Delta^E} \right] 
\end{eqnarray}
and
\begin{equation}
\zeta^2(\varepsilon-1)
= 4\kappa^2\left( \frac{1}{s}-1 \right) \frac{1}{s},
\end{equation}
which brings out the divergence structure and the related cancellations
in the perfect conductor limit.
Thus, the contribution from the TE-mode in Eq.\,(\ref{trgepEv}) is
\begin{equation}
\text{TE}: \quad -\zeta^2(\varepsilon-1)
g^E_{\rput(0.1,0){\scriptstyle{4}} \pscircle[linewidth=0.2pt](0.1,0.01){0.13}}
(a_1,a_1) = 2\kappa - \frac{2\kappa}{s}
+2\kappa \frac{r_\varepsilon \bar r_\mu}{\Delta^E} e^{-2\kappa a}.
\label{temoc}
\end{equation}
The rest of the contributions in Eq.\,(\ref{trgepEv}) is the TM-mode,
which can be shown to be
\begin{eqnarray}
\text{TM}: \quad \frac{1}{\varepsilon} \partial \partial^\prime
g^H_{\rput(0.1,0){\scriptstyle{4}} \pscircle[linewidth=0.2pt](0.1,0.01){0.13}}
(a_1,a_1) +\frac{k_\perp^2}{\varepsilon}
g^H_{\rput(0.1,0){\scriptstyle{4}} \pscircle[linewidth=0.2pt](0.1,0.01){0.13}}
(a_1,a_1)  \nonumber \hspace{25mm} \\
= 2\kappa - \frac{2\kappa}{s}
+2\kappa \frac{\bar r_\varepsilon r_\mu}{\Delta^H} e^{-2\kappa a}.
\hspace{15mm}
\label{tmmoc}
\end{eqnarray}
The evaluation of the contribution to the TM mode was assisted
 by the identities~\cite{Prachi:2011ec}
\begin{equation}
k_\perp^2 \pm \kappa\kappa_\varepsilon 
= -\frac{\bar\kappa_\varepsilon \mp\kappa}{\kappa_\varepsilon \mp\kappa}
\zeta^2 \varepsilon.
\end{equation}
Using the contribution from TE mode in Eqs.\,(\ref{temoc})
and TM mode in Eq.\,(\ref{tmmoc})
to evaluate Eq.\,(\ref{trgepEv}), and using it in Eq.\,(\ref{cfintg}),
and rewriting $2\kappa/s =\kappa_\varepsilon +\kappa$,
the total pressure in Eq.\,(\ref{cfintg}) takes the form
\begin{eqnarray}
\frac{{\bf F} \cdot \hat{\bf z}}{A}
&=& \int_{-\infty}^\infty \frac{d\zeta}{2\pi}
\int \frac{d^2k_\perp}{(2\pi)^2} 
\nonumber \\ && \times
\left[ \kappa -\kappa_\varepsilon
+\kappa \left( \frac{r_\varepsilon \bar r_\mu}{\Delta^E}
+\frac{\bar r_\varepsilon r_\mu}{\Delta^H} \right) e^{-2\kappa a} \right],
\hspace{8mm}
\label{boy-finexp}
\end{eqnarray}
where $\Delta^E$ and $\Delta^H$ are given using Eqs.\,(\ref{delEH}).
The first term in Eq.\,(\ref{boy-finexp}) is the contribution from
the empty space when the slabs are removed, which is divergent.
The second term is the contribution
due to the change in the single-body energy due to the variation in 
the dielectric medium, which is also divergent.
The remaining term in the perfect conductor limit leads to 
\begin{equation}
\frac{{\bf F} \cdot \hat{\bf z}}{A}
= -\frac{1}{\pi^2} \int_0^\infty \frac{\kappa^3 d\kappa}{e^{2\kappa a}+1}
=-\frac{7}{8} \frac{\pi^2}{240 a^4}.
\label{boy-fpcexp}
\end{equation}
Again, we reproduce the Boyer result.

%--------------------------------------
\subsection{Juxtaposition}

In the Casimir (or Lifshitz) configuration
of Fig.~\ref{fig-casimir-config} when we require both slabs
to have the same dielectric property, $\varepsilon >1$,
with no magnetic property, $\mu=1$, we recall, for example in 
Ref.~\cite{Schwinger:1978dec}, the corresponding expression for the
Casimir force to be
\begin{eqnarray}
\frac{{\bf F}_\text{Cas} \cdot \hat{\bf z}}{A}
&=& \int_{-\infty}^\infty \frac{d\zeta}{2\pi}
\int \frac{d^2k_\perp}{(2\pi)^2} 
\nonumber \\ && \times
\left[ \kappa -\kappa_\varepsilon
+\kappa \left( \frac{r_\varepsilon^2}{\Delta^E_\text{Cas}}
+\frac{\bar r_\varepsilon^2}{\Delta^H_\text{Cas}} 
\right) e^{-2\kappa a} \right], \hspace{9mm}
\label{cas-finexp}
\end{eqnarray}
where
\begin{subequations}
\begin{eqnarray}
\Delta^E_\text{Cas} &=& 1 -r_\varepsilon^2 e^{-2\kappa a}, \\
\Delta^H_\text{Cas} &=& 1 -\bar r_\varepsilon^2 e^{-2\kappa a}.
\end{eqnarray}
\end{subequations}
In the perfect conductor limit the expression in
Eq.\,(\ref{cas-finexp}) takes the form
\begin{equation}
\frac{{\bf F}_\text{Cas} \cdot \hat{\bf z}}{A}
= \frac{1}{\pi^2} \int_0^\infty \frac{\kappa^3 d\kappa}{e^{2\kappa a}-1}
= \frac{\pi^2}{240 a^4},
\label{cas-fpcexp}
\end{equation}
which is the attractive Casimir pressure on the left perfect conducting slab.

The display of Eq.\,(\ref{boy-finexp}), that leads to the Boyer force,
and Eq.\,(\ref{cas-finexp}), that leads to the Casimir force, 
brings out the close similarity in the two expressions, and
reveals the fine differences between them. We observe that the
difference in the reflection coefficients of the second slab
in the Boyer and Casimir configurations in Figs.~\ref{fig-boyer-config} 
and \ref{fig-casimir-config} is the source of the dissimilarity
in the two expressions.
We note that the relative sign difference between the Boyer
and the Casimir configuration arises because
in the perfect electric conducting limit ($\varepsilon\to\infty$)
and perfect magnetic conducting limit ($\mu\to\infty$) we obtain
a relative sign difference, 
\begin{equation}
r_\varepsilon \bar r_\mu \to -1, \quad
\text{and} \quad \bar r_\varepsilon r_\mu \to -1, 
\end{equation}
versus 
\begin{equation}
r_\varepsilon^2 \to 1, \quad \text{and} \quad
\bar r_\varepsilon^2 \to 1.
\end{equation}
Thus, in perfect conducting limits,
after dropping the first two divergent terms in 
Eqs.\,(\ref{boy-finexp}) and (\ref{cas-finexp}),
that removes the bulk and single-body contributions,
the expression for the pressure on the left slab is given by the
third term in Eqs.(\ref{boy-finexp}) and (\ref{cas-finexp}),
which differ by an overall sign, in addition to the difference in
sign in their denominators. This is the difference in 
sign (direction) between the pressures on the left slab in
the Boyer and Casimir configuration. We also note that,
in the perfect conducting limits, $\Delta^E$ and $\Delta^H$ for the
Boyer configuration go to $1+e^{-2\kappa a}$, while the corresponding
factors $\Delta^E_\text{Cas}$ and $\Delta^H_\text{Cas}$
in the Casimir configuration go to $1-e^{-2\kappa a}$.
These renders to the difference in sign in the denominators
of Eqs.\,(\ref{boy-fpcexp}) and (\ref{cas-fpcexp}),
and leads to the classic relative factor of
\begin{equation}
\frac{7}{8} = \dfrac{ 
{\displaystyle \int_0^\infty} \dfrac{\kappa^3 d\kappa}{e^{2\kappa a}+1} }{
{\displaystyle \int_0^\infty} \dfrac{\kappa^3 d\kappa}{e^{2\kappa a}-1} }
\end{equation}
in Eq.\,(\ref{boy-fpcexp}).

The Boyer configuration and the Casimir configuration
of Figs.~\ref{fig-boyer-config} and \ref{fig-casimir-config}
can, both, independently, be interpreted to form a cavity,
which are by construction different in their constituent boundaries.
We shall call them the Boyer cavity and the Casimir cavity.
If a monochromatic electromagnetic wave completes one closed loop 
in a Casimir cavity, involving a reflection off the right dielectric slab
and then a reflection off the left dielectric slab, it is effectively
unchanged, because $r_\varepsilon^2 =1$. On the other hand, if
a monochromatic electromagnetic wave completes a closed loop in
a Boyer cavity, it develops a phase difference of $\pi$ because 
$r_\varepsilon \bar r_\mu =-1$. Further, after completing two
closed loops in a Boyer cavity the wave returns back to its
original state. In this sense, the Boyer cavity is topologically
different from a Casimir cavity.
 
In the multiple scattering formalism,
the Casimir force and the Boyer force can be interpreted as the sum
of all possible scattering inside the respective cavities. That is, 
the integrals in Eqs.\,(\ref{boy-fpcexp}) and (\ref{cas-fpcexp}),
respectively, are interpreted as the sum of all possible
permutations and combinations of scattering possible in the
respective cavities. This can be more explicitly illustrated by
expanding the integrand of 
Eqs.\,(\ref{boy-fpcexp}) and (\ref{cas-fpcexp}),
as a binomial expansion in $e^{-2\kappa a}$.
This assemblage of all possible permutations and combinations
of scattering inside a cavity gives the cavity itself a
statistical identity. 
In other words, the spectral distribution of scattering inside
a Casimir cavity is described by Bose-Einstein distribution,
while the spectral distribution of scattering inside
a Boyer cavity is described by Fermi-Dirac distribution.
Thus, in conclusion, the source for the difference in
the Casimir force and the Boyer force lies in the fact that
the respective cavities have fundamentally different statistics
describing the scattering inside the respective cavities.

The Boyer configuration and the Casimir configuration in 
Figs.~\ref{fig-boyer-config} and \ref{fig-casimir-config}
are special cases of more general boundary conditions that has
been studied for two magneto-electric $\delta$-function 
plates in Ref.~\cite{Milton:2013bm}.
Similar general boundary conditions for a scalar field has been
extensively studied in Ref.~\cite{Asorey:2013wca}.
This, then, suggests a class of statistics associated
with the distribution of scatterings inside such general configurations.
However, these characteristics are dependent on the parameters
describing the material. The special extreme limits of Boyer and
Casimir configuration has the nice feature that it has no dependence
on the particular material properties.

%--------------------------------------------
\section{Finite coupling using stress tensor method}
\label{sec-finitestm}

The perfect coupling limit in Sec.~\ref{sec-per-con-limit}
and Sec.~\ref{sc} is an extreme limit, and it is instructive 
to consider the finite coupling case for real materials.
We need to evaluate the pressure along the $z$-direction given
in (\ref{P-a1}) using the scalar electric and magnetic 
Green's functions for the finite value of permittivity
and permeability in Eqs.~(\ref{t33-}) and (\ref{t33+}).
We obtain
\begin{equation}
T_{33}(a_1-\delta) = -\frac{\kappa_\varepsilon}{i},
\label{t33-f}
\end{equation}
which is just the contribution coming from the single-body bulk terms in
the Green function. While 
\begin{equation}
T_{33}(a_1+\delta) = -\frac{\kappa}{i} -\frac{\kappa}{i}
\bigg( \frac{r_\varepsilon \bar{r}_\mu}{\Delta^E}
+\frac{\bar{r}_\varepsilon r_\mu} {\Delta^H}  \bigg)e^{-2\kappa a}.
\label{t33+f}
\end{equation}
Dropping the bulk contributions in 
$T_{33}(a_1-\delta)$ and $T_{33}(a_1+\delta)$ in 
Eqs.\,(\ref{t33-f}) and (\ref{t33+f}) we find
\begin{equation}
P_\varepsilon = \int_{-\infty}^{\infty} \frac{d\zeta}{2\pi}\int 
\frac{d^2k_\perp}{(2\pi)^2}
\kappa\bigg( \frac{r_\varepsilon \bar{r}_\mu}{\Delta^E}
+\frac{\bar{r}_\varepsilon r_\mu}{\Delta^H}  \bigg)e^{-2\kappa a},
\label{real-P}
\end{equation}
which is identical to the what we obtained in Eq.\,(\ref{boy-finexp}),
if the dropped terms were reintroduced.
We can write Eq.~(\ref{real-P}) in spherical polar co-ordinate
system after defining $k_\perp=\kappa \sin\theta$ and 
$\zeta=\kappa \cos\theta$, then it is straightforward
to carry out the $\kappa$ integration, to obtain
\begin{equation}
P_\varepsilon = \frac{3}{16\pi^2 a^4}\int_0^1 d\,t
\left[ \text{Li}_4(r_\varepsilon \bar{r}_\mu)
+ \text{Li}_4(\bar{r}_\varepsilon r_\mu)\right],
\end{equation}
where $\text{Li}_4(x)$ is the polylogarithm function and
\begin{subequations}
\begin{eqnarray}
r_\varepsilon(t) &=& \frac{\sqrt{1+(\varepsilon-1) t^2}-1}
{\sqrt{1+(\varepsilon-1) t^2}+1}, \\
r_\mu(t) &=& \frac{\sqrt{1+(\mu-1) t^2}-1}{\sqrt{1+(\mu-1) t^2}+1},
\end{eqnarray}
\end{subequations}
and
\begin{subequations}
\begin{eqnarray}
\bar r_\varepsilon(t) &=& \frac{\sqrt{1+(\varepsilon-1) t^2}-\varepsilon}
{\sqrt{1+(\varepsilon-1) t^2}+\varepsilon}, \\
\bar r_\mu(t) &=& \frac{\sqrt{1+(\mu-1) t^2}-\mu}{\sqrt{1+(\mu-1) t^2}+\mu}.
\end{eqnarray}
\end{subequations}
It is not a priori obvious whether this pressure is
attractive or repulsive on the left boundary. Even for the simple
non-dispersive case it is not easy to do the remaining 
integration analytically. The plot of the integrand from 0 to 1
is always negative, implying that the pressure on the left boundary
is still repulsive. We can carry out the integration numerically
and observe that the ratio of the pressure on left wall for 
the real material to the Boyer's pressure is always positive
and approaches Boyer's result for very large values of permittivity
and permeability.

To make our analysis include dispersion, we resolve to a simple
model for the permeability as described in Eq.~(\ref{11}) in
which we breakup the frequency integration in 
(\ref{real-P}) into two parts. We assume $\mu$ to be constant
up to a critical frequency $\zeta_c=-i\omega_c$, thereafter we set $\mu=1$.
Clearly, we will not get any contribution for the 
higher frequencies beyond $\zeta_c$ for our system as there is
no interface on the right when $\mu=1$. 

\begin{figure}
\includegraphics[width=8cm]
{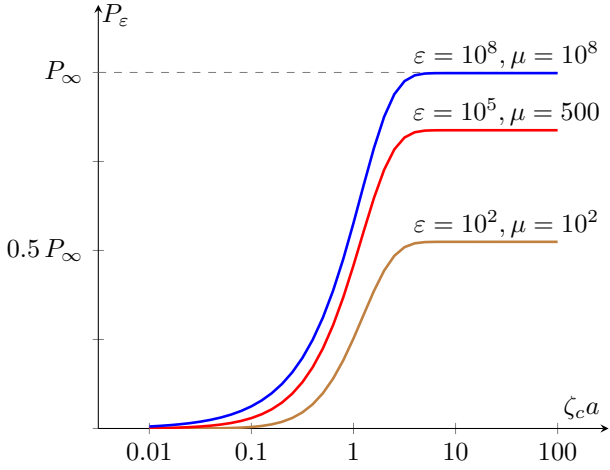}
\caption{The Casimir pressure on the left interface of a material
medium with electrical permittivity $\varepsilon$ interacting
with a material medium of magnetic permeability $\mu$ with
cutoff frequency $\zeta_c$ in Eq.\,(\ref{11}) plotted as a 
function of $(\zeta_c a)$.
Here $P_\infty$ is the Boyer result for the perfectly electrically
conducting and perfectly magnetically conducting half-spaces.
Realisitic frequency responses in magnetic materials do not 
go beyond 1\,MHz, which corresponds to $\zeta_c a\sim 10^{-8}$
for $a=1\,\mu$m, for which the pressure is immeasurably small.
}
\label{PoverP0-0-5}
\end{figure}

In Fig.~\ref{PoverP0-0-5}, we plot the pressure $P_\varepsilon$
normalized to the Boyer value for the perfectly conducting
case $P_\infty$ as a function of the dimensionless product 
$\zeta_ca$ in Eq.\,(\ref{11}) for different values of permittivity
and permeability. For reference, at $a=1\,\mu$m the cutoff 
frequency is $\omega_c=c/a=3\times 10^{14}$\,Hz.
The positive values for $P_\varepsilon/P_\infty$
implies that the pressure between a real electric and magnetic
media is still repulsive. For very high values of 
respective permittivity and permeability of two media,
the Casimir pressure approaches Boyer's result as expected,
while for more realistic values it remains less than Boyer's result.
Realisitic frequency responses in magnetic materials do not  
go beyond 1\,MHz, which corresponds to $\zeta_c a\sim 10^{-8}$
for $a=1\,\mu$m, for which the pressure is immeasurably small.
This connects with the fact that the Boyer repulsion
has never been experimentally measured.

%---
%---

%--------------------------------------------
\section{Statistical mechanical considerations}
\label{secintro}

In this section we try to give an alternative explanation of
the Boyer problem by drawing quantum statistical mechanics
into consideration.

Let the two media (half-spaces),  separated by a gap $a$,
be general dielectrics endowed with arbitrary constant values of
$\varepsilon$ and $\mu$. We introduce a model where the media are
represented by  harmonic oscillators 1 and 2,  interacting with
each other via a third oscillator 3. The last oscillator represents
the electromagnetic field. In the generalized version of the model,
the third oscillator represents an assembly of oscillators. An advantage
of this model is that it provides a mechanical analogue to the
conventional TE and TM modes in electromagnetism. Among these,
the TM modes are the easiest ones to visualize, among else things
because the induced interaction increases when the temperature $T$
increases, so as to reach the classical limit for high $T$. By contrast,
in the TE case the force goes to zero in the classical case because
the zero Matsubara frequency will not contribute.
And, as we will show, this harmonic oscillator model is able to
shed some further light on the Boyer problem also. We have actually
introduced the model in earlier works, in connection with the
Casimir effect, \cite{hoye03,hoye16,hoye18},
although the model does not seem to be well known.

Let us sketch  some essentials of the formalism. As is known,
the classical partition function $Z$ of a harmonic oscillator is
$Z=1/(\hbar \beta \omega_i)$, with $\beta=1/(k_BT)$,
where $\omega_i$ here is the eigen frequency of an oscillator.
It means that the free energy is
 \begin{equation}
 F=-\frac{1}{\beta}\ln Z \sim \ln \omega_i.
 \end{equation}
If three noninteracting oscillators the inverse partition function
is thus proportional to $\sqrt{Q}$, where
\begin{equation}
Q=a_1 a_2 a_3, \quad a_i=\omega_i^2 \quad (i=1,2,3).
\label{}
\end{equation}
When going over to quantum statistical mechanics
(path integral method \cite{hoye81,brevik88}),
the classical system is imagined to be divided into a set of
classical harmonic oscillator systems. It implies the substitutions
\begin{equation}
Q=A_1 A_2 A_3, \quad A_i=a_i+\zeta^2 = \omega_i^2+\zeta^2
\label{A}
\end{equation}
where now $\zeta=i\omega$ is the Matsubara frequency.

We assume that oscillators 1 and 2 interact via oscillator 3,
and assume for simplicity that all oscillators are one-dimensional.
Taking the interaction to be bilinear, thus expressible in the form
$cx_ix_j$ with $c$ a coupling constant, we obtain the following determinant
\begin{subequations}
\begin{eqnarray}
Q=\left|
\begin{array}{ccc}
A_1 & 0 & c\\
0 & A_2 & c\\
c & c & A_3\\
\end{array}
\right| &=& %\\ && 
A_1 A_2 A_3(1-D_1)(1-D_2) \nonumber \\
&& \times\left[1-\frac{D_1 D_2}{(1-D_1)(1-D_2)}\right], \hspace{8mm}
\label{Q}
\end{eqnarray}
\end{subequations}
where
\begin{equation}
D_j=\frac{c^2}{A_j A_3}\quad (j=1,2).
\end{equation}
Here the first term $A_1A_2A_3$ refers to the noninteracting oscillators.
The terms $A_j(1-D_j)\, (j=1,2)$ are the separate interactions between
1-3 and 2-3, while the last term is the Casimir energy.
If $Q>0$, the Casimir energy is negative, corresponding
to an attractive force.

This is the straightforward part of the analysis,
and corresponds to the electromagnetic TM mode. The TE mode is more
intricate, as it  corresponds to oscillator 3 interacting with the
{\it momenta} of oscillators 1 and 2. The interaction replaces
the term $p_j^2/2m_j$ with $({\bf p}_j-e{\bf A})^2/2m_j$,
where ${\bf{p}}_j$ is the canonical momentum
and $\bf A$ the vector potential. In the oscillator model
the analogous energy takes the form
\begin{equation}
 \frac{1}{2}m_ja_j\left(p_j-\frac{c}{a_j} x_3\right)^2 \label{E}
\end{equation}
Now again  $c$ is the coupling parameter and $x_3$ corresponds to $A_3$.

The important outcome of this is that the $A_3$ in the determinant $Q$
is changed,
\begin{equation}
A_3\rightarrow A_3+\frac{c^2}{a_1}+\frac{c^2}{a_2}. \label{B}
\end{equation}
This means that $Q$ can be written as 
\begin{equation}
Q=\left| \begin{array}{ccc}
A_1 & 0 & c\\ 0 & A_2 & c\\ cq_1 & cq_2 & A_3\\
\end{array} \right| 
= \left| \begin{array}{ccc}
A_1 & 0 & \frac{\zeta c}{\sqrt{a_1}}\\ 0 & A_2 & \frac{\zeta c}{\sqrt{a_2}}\\
-\frac{\zeta c}{\sqrt{a_1}} &-\frac{\zeta c}{\sqrt{a_2}}  & A_3\\
\end{array} \right|, 
\label{Q1}
\end{equation}
using the property of determinants in the second equality, where
\begin{equation}
q_j=1-\frac{A_j}{a_j}=-\frac{\zeta^2}{a_j}, \quad(j=1,2).
\label{}
\end{equation}
The determinant $Q$ can still be expressed  as in Eq.~(\ref{Q}), but now with 
\begin{equation}
D_j=-\frac{\zeta^2 c^2}{a_j A_j A_3}<0.
\end{equation}
The induced force is still attractive since $D_1 D_2>0$,
although both factors are negative. 

Proceed now to the third and final step. Assume for definiteness
that only oscillator $j=2$ has the form (\ref{E}). Then $A_3$ changes to
\begin{equation}
A_3\rightarrow A_3+\frac{c^2}{a_2},
\label{}
\end{equation}
and the determinant (\ref{Q1}) changes to
\begin{eqnarray}
Q=\left|
\begin{array}{ccc}
A_1 & 0 &  c\\
0 & A_2 & \frac{\zeta c}{\sqrt{a_2}}\\
c &-\frac{\zeta c}{\sqrt{a_2}}  & A_3\\
\end{array}
\right|.
\end{eqnarray}
This implies that
\begin{equation}
D_1=\frac{c^2}{A_1 A_3}>0 \quad \mbox{and} 
\quad D_2=-\frac{\zeta^2 c^2}{a_j A_j A_3}<0,
\end{equation}
so that $D_1 D_2<0$ and the induced force becomes repulsive. 
 
Looking back, we are now able to summarize what were the
characteristic properties of the harmonic oscillator model making the
transition from an attractive to a repulsive force possible:
 
\begin{enumerate}
\item The generalization $a_i \rightarrow A_i$ in Eq.~(\ref{A}),
meaning incorporation of the discrete Matsubara frequencies $\zeta$;
\item the modification of $A_3$ in Eq.~(\ref{B}), corresponding
to the transition from the TM to the TE mode. In turn, this is
related to the oscillator 3 interacting with oscillators 1 and 2
via canonical momenta $p_j$, instead of via  positions $x_i$
(interaction energy  $cx_ix_j$) as was characteristic for the TM mode.
\end{enumerate}
 
When seen in this way, the basic reason for the Boyer problem is
linked to quantum mechanics. This is a satisfactory conclusion,
since otherwise, in classical electrodynamics, the sign reversal
of $E^2$ in the force expression 
$-\frac{1}{2} E^2{\bm\nabla} \varepsilon$
as noted above, would be quite non-understandable.

%-----------------------------------------------------
\acknowledgments

We acknowledge support from the Research Council of Norway
(Project No. 250346).

%-----------------------------------------------------
\bibliography{biblio/b20170713-gdbc}
%\nocite{*} %%% Will print the complete bib data.
%-----------------------------------------------------

\end{document}